\documentclass[10pt, a4paper, twocolumn, aps, pra, showpacs, longbibliography, nofootinbib,superscriptaddress]{revtex4-1}
\pdfoutput=1
\usepackage[utf8x]{inputenc}
\usepackage{ucs}
\usepackage{amsmath}
\usepackage{amsfonts}
\usepackage{amssymb}
\usepackage{makeidx}
\usepackage{cellspace,booktabs}
\usepackage{natbib}
\usepackage{lipsum}
\usepackage{bm}
\usepackage{bbm}
\usepackage{dsfont}
\usepackage{relsize}

\usepackage{hyperref}
\hypersetup{colorlinks = true, linkcolor=magenta,citecolor=blue, urlcolor=magenta, bookmarksnumbered =  true}

\usepackage{color}
\definecolor{light-gray}{gray}{0.55}

\usepackage{microtype}

\usepackage{graphicx}

\newcommand{\ssm}{\rm\scriptscriptstyle}

\newcommand{\ket}[1]{ \lvert #1 \rangle}

\begin{document}

\begin{abstract}	
Machine-learning driven models have proven to be powerful tools for the identification of phases of matter. In particular, unsupervised methods hold the promise to help discover new phases of matter without the need for any prior theoretical knowledge. While for phases characterized by a broken symmetry, the use of unsupervised methods has proven to be successful, topological phases without a local order parameter seem to be much harder to identify without supervision. Here, we use an unsupervised approach to identify topological phases and transitions out of them. We train artificial neural nets to relate configurational data or measurement outcomes to quantities like temperature or tuning parameters in the Hamiltonian. The accuracy of these predictive models can then serve as an indicator for phase transitions. We successfully illustrate this approach on both the classical Ising gauge theory as well as on the quantum ground state of a generalized toric code.
\end{abstract}

\date{\today}
\author{Eliska Greplova}
\affiliation{Institute for Theoretical Physics, ETH Zurich, CH-8093, Switzerland}
\author{Agnes Valenti}
\affiliation{Institute for Theoretical Physics, ETH Zurich, CH-8093, Switzerland}
\author{Gregor Boschung}
\affiliation{Institute for Theoretical Physics, ETH Zurich, CH-8093, Switzerland}
\author{Frank Schäfer}
\affiliation{Department of Physics, University of Basel, Klingelbergstrasse 82, CH-4056 Basel, Switzerland}
\author{Niels Lörch}
\affiliation{Department of Physics, University of Basel, Klingelbergstrasse 82, CH-4056 Basel, Switzerland}
\author{Sebastian Huber}
\affiliation{Institute for Theoretical Physics, ETH Zurich, CH-8093, Switzerland}

\title{Unsupervised identification of topological order using predictive models}

\maketitle

\section{Introduction}

Identifying phase transitions is one of the key questions in theoretical and experimental condensed matter physics alike. For the experimental characterization of thermodynamic phase transitions, there exists an excessive amount of possible tools, ranging from system specific, like the study of the conductivity in an electronic system, to very generic, like the specific heat. The latter is particularly appealing as it does not assume any prior knowledge: For example, structural transitions, the onset of magnetism, or the transition to superconductivity, all show up in this generic probe. The study of the specific heat is also a standard tool for the theoretician, especially given its generic power. 

For quantum phase transitions \cite{Sachdev2011}, an equally generic tool as the specific heat for thermal transitions is the fidelity susceptibility. One investigates the derivative of the overlap $\partial_\beta \langle\psi(\beta+\epsilon)|\psi(\beta)\rangle$ \cite{YouLi2007} of two infinitesimally separated ground states $|\psi(\beta)\rangle$ as a function of some tuning parameter $\beta$. While this probe is in principle very powerful \cite{VenutiZanardi2007, GuLin2009, ZanardiPaunkovic2006, AbastoHamma2008}, it is typically hard to evaluate as one has rarely access to the full wave-function. At least not for most of the approximate numerical techniques and especially not in experimental studies. This raises the question if one can replace the fidelity susceptibility with a tool that is equally {\em unbiased}, {\em generic}, and {\em accessible} to typical numerical and experimental techniques.

In a recent publication some of the present authors introduced such an algorithmic method for classical systems with an order-parameter signaling an (arbitrary) symmetry breaking \cite{Schafer2019}. Here we demonstrate that one can successfully generalize this method to problems without a local order parameter, i.e., systems with a topological character. Moreover, we show that one can straightforwardly extend Ref.~\onlinecite{Schafer2019} to the quantum realm. 

The method is based on the analysis of the accuracy of a predictive model. The central idea is to distill a predictive model that relates input data from numerical or experimental studies to the output in the form of a known tuning parameter such as the temperature or a parameter in the Hamiltonian $\beta$. Typically, one infers this predictive model via machine-learning techniques in the form of neural nets. The basic idea, however, is independent of the specific inference technique. In a next step, the accuracy of the predictive model is analyzed via the comparison of the predicted to the known value of the tuning parameter $\beta$. In particular, we show the derivative of the prediction accuracy with respect to the tuning parameter to be an equally sensitive indicator of a phase transition as the fidelity susceptibility.

To illustrate our generalization of the methods of Ref.~\onlinecite{Schafer2019}, we investigate two generic models hosting interesting thermodynamic phases without a local order parameter. First, we investigate the finite-temperature cross-over in Wegner's Ising gauge theory (IGT) \cite{Wegner1971, FradkinSusskind1978, Kogut1979} to show that we can analyze an interesting classical problem without a local order parameter. Second, we broaden the scope by taking the step from the IGT to a generalized toric code problem \cite{Kitaev2006, CastelnovoChamon2007} showcasing the applicability of the method to quantum problems.

\section{The Ising gauge theory}
\label{sec:IGT}

\begin{figure}
\centering
\includegraphics[scale=0.32]{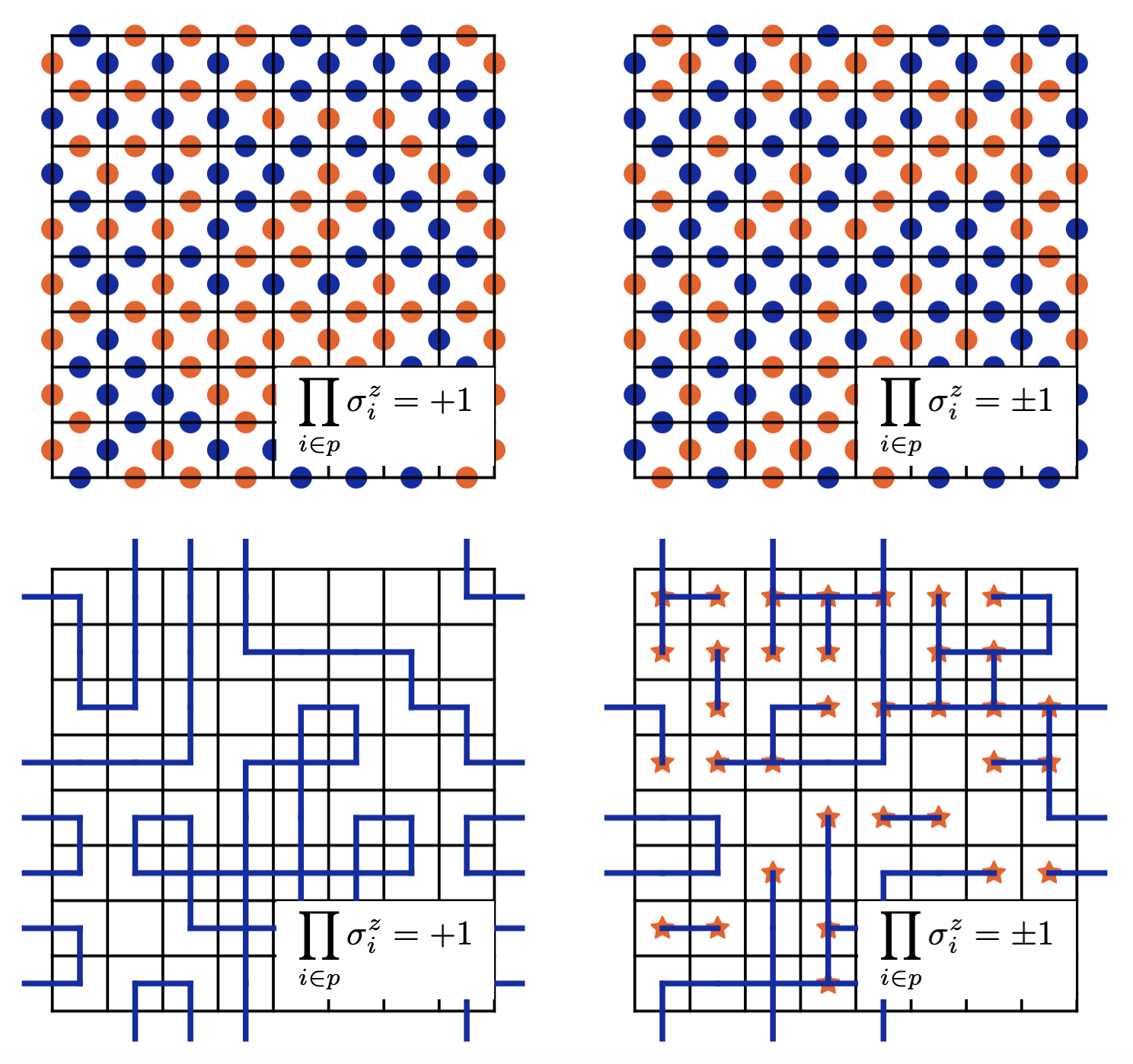}
\caption{Local constraint of Ising gauge theory: upper left panel shows an example of $T=0$ state where the gauge condition $\prod_{i}\sigma_i^z=1$ is met for all plaquettes, lower left panel shows the corresponding dual map, where spins are mapped on Wilson loops that are uninterrupted in the case of zero-temperature states. Upper right panel shows an example of a $T=\infty$ state, where the gauge condition is violated, lower panel shows corresponding Wilson loops with breakages at the places the plaquette condition is not met.}
\label{fig:igt_example}
\end{figure}

Wegner's Ising gauge theory (IGT) is a spin model defined on a $N\times N$ square lattice with spins placed on the lattice bonds \cite{Wegner1971, FradkinSusskind1978, Kogut1979, Sachdev2018}. It is described by the Hamiltonian
\begin{equation}
	\label{eqn:igt}
H_{\ssm IGT} = - J \sum_{p}\prod_{i\in p}\sigma_i^z,
\end{equation}
where $J$ is a coupling constant, $p$ refers to plaquettes on the lattice (see Fig.~\ref{fig:igt_example}), and $\sigma_i^z$ is the Pauli matrix describing a single spin-1/2. The ground state of this Hamiltonian is a highly degenerate manifold, an arbitrary superposition of all states that meet the condition that the product of spins along each plaquette is equal to $1$. At a finite temperature $T>0$ the local constraints $\prod_{i\in p}\sigma_i^z=1$ are violated (see Fig.~\ref{fig:igt_example}). The IGT does not have a finite temperature phase transition. However, for finite system sizes one can find a crossover temperature, $T^*=1/(k\beta^*)$ defined by the appearance of one plaquette with $\prod_{i\in p}\sigma_i^z=-1$, resulting in the scaling $T^*\sim 1/\ln (2N^2)$ \cite{CastelnovoChamon2007, CarrasquillaMelko2017}. Matters are further complicated by the fact that the ground-state manifold cannot be characterized by a local order parameter \cite{Elitzur1975, Kogut1979} owing to a local gauge degree of freedom. We come back to this point below.

To check whether a given spin state is in the IGT ground-state manifold, one has to verify that the condition $\prod_{i\in p}\sigma_i^z=1$ is met for all plaquettes in the lattice. Equivalently, one can use the duality map to analyze the phase transition: We  connect the edges of the lattice that contain spins with the same orientation and form loops. The IGT constrained phase then has the property that all these loops are closed. Whenever the constraint is violated it results in an open loop \cite{Kogut1979, Giles1981, Sachdev2018}, see Fig.~\ref{fig:igt_example}.

Distinguishing high and low temperature states of the model (\ref{eqn:igt}) is a well studied test case for machine learning recognition of phases of matter \cite{CarrasquillaMelko2017}. As one can see from Fig.~\ref{fig:igt_example}, the IGT constitutes an interesting example where the phases are hard to distinguish visually without being a-priori familiar with a local restrictions or the dual map. While an supervised approach is immediately successful at distinguishing the high and low temperature phases \cite{CarrasquillaMelko2017}, unsupervised approaches did not succeed without an explicit recipe what type of restriction to look at. There has been significant progress in this direction, but a fully general approach is yet to be found \cite{Rodriguez2018, Zhao2019}. While methods like principal component analysis, clustering and variational auto-encoders have proven to be successful to determine the phase transitions in spin models possessing an order parameter \cite{Wetzel2017}, systems without order parameters still represent a challenge. 

Here we show how the method introduced by Sch\"afer et al. \cite{Schafer2019} can be generalized to systems without a local order parameter. One first pre-trains a neural network to relate a spin configuration $\{S\}_{\beta_{\ssm label}}$ to the (inverse) temperature $\beta_{\ssm label}$, at which the configuration was sampled. After this initial training, the performance of the estimator is assessed with respect to the true value. The derivative \begin{equation}
\mathcal D(\beta_{\ssm label})=\frac{\partial}{\partial \beta_{\ssm label}} \beta_{\ssm pred}(\{S\}_{\beta_{\ssm label}})
\label{eq:notdivergence}
\end{equation}
is maximal where the estimator performs worst. In other words, a local maxima in $\mathcal D(\beta_{\ssm label})$ indicate a phase transition or cross-over temperature $\beta_{\ssm label}^*$. While this method does not in principle rely on a local order parameter, it has been show that the network picks up on the magnetization pattern \cite{Schafer2019}. It was therefore unclear if one can generalize this strategy to the current problem. Here we show that this approach is valid even for phases of matter that do not contain an order parameter, or a finite temperature phase transition. 

We create sample configurations of the IGT model and label them with $\beta=1/(kT)$. We train a convolutional neural network to predict $\beta$ given an IGT configuration as an input. Our neural network consists of $2$ convolutional and $2$ dense layers and was trained on $2\cdot 10^5$ configurations for $100$ different values of $\beta$ (for details see Appendix~\ref{app:igt}). 

\begin{figure}
\centering
\includegraphics[scale=1]{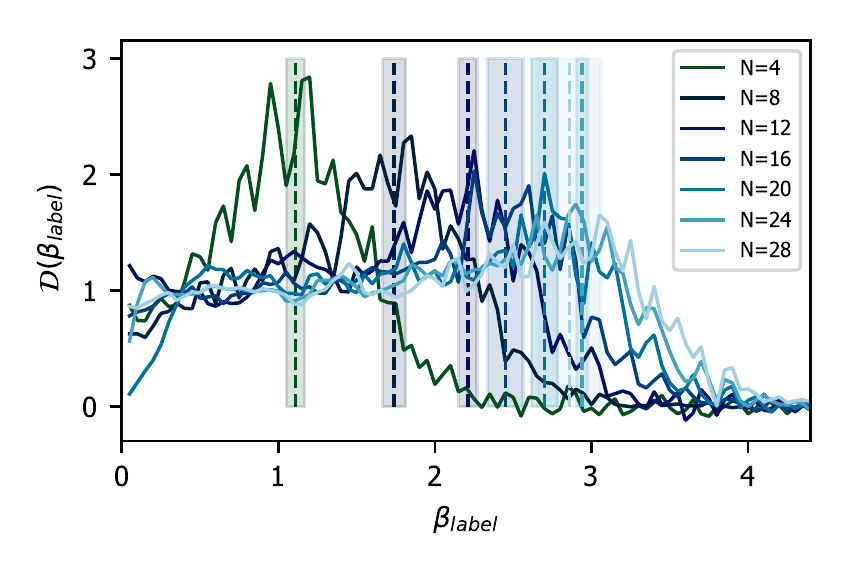}
\caption{Derivative of the output of the predictive model, $\mathcal D(\beta_{\ssm label})$, as a function of assigned labels $\beta_{label}$ for system sizes $N=4$ to $N=28$. The dashed lines denote the position of the crossover inverse temperature $\beta^*$ as determined by the density of states method.}
\label{fig:divergence_igt}
\end{figure}

In Fig.~\ref{fig:scaling_igt} we show how the difference between the true and predicted inverse temperatures  $\beta_{\ssm pred}-\beta_{\ssm label}$ behaves as a function of the true $\beta_{\ssm label}$ for seven different system sizes $N = 4, 8, 12, 16, 20, 24, 28$ (the total number of spins is $2N^2$). We see that the behavior of the prediction is not uniform for all inputs and, in fact, we observe that for all systems sizes there exist different finite $\bar \beta$ above which the network has difficulties to identify the correct $\beta_{\ssm label}$. In Fig.~\ref{fig:divergence_igt} we show $\mathcal D(\beta_{\ssm label})$ which we evaluated as 
\begin{align}
\mathcal D(\beta_{\ssm label})\approx
\frac{\beta_{\ssm pred}\left(\{S\}_{\beta^{i+1}_{\ssm label}}\right)-\beta_{\ssm pred}\left(\{S\}_{\beta^{i-1}_{\ssm label}}\right)}{\beta^{i+1}_{\ssm label}-\beta^{i-1}_{\ssm label}} \nonumber,
\end{align}
where sampled at discrete $\beta^{i}_{\ssm label}$. For all system sizes we observe a presence of a peak that indicates the position of the largest change in the difference between true and predicted $\beta$.

\begin{figure}
\centering
\includegraphics[scale=1]{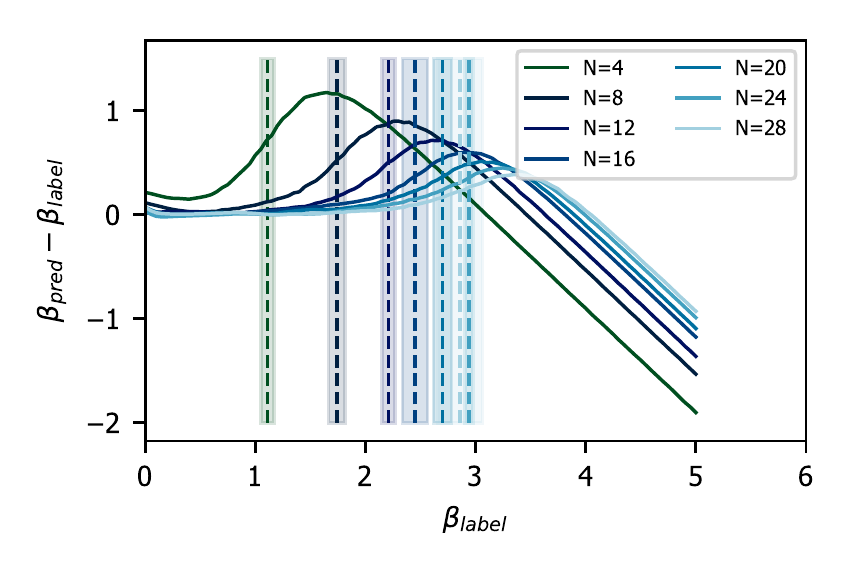}
\caption{We show the difference of the network prediction of $\beta_{\ssm pred}$ and assigned label $\beta_{\ssm label}$, $\beta_{\ssm pred} - \beta_{\ssm label}$ as a function of $\beta_{label}$ for system sizes $N=4$ to $N=28$. The dashed lines denote the position of the crossover inverse temperature $\beta^*$ as determined by the density of states method.}
\label{fig:prediction_igt}
\end{figure}

The neural network predicts a continuous parameter (inverse temperature) for our model and we observe a change of behavior at some critical value. We show in Fig.~\ref{fig:scaling_igt} the determined crossover temperature $\beta^*$ as a function of system size. For the system sizes we were able to test numerically we recover logarithmic scaling as expected for the crossover temperature \cite{CarrasquillaMelko2017, CastelnovoChamon2008}.

\begin{figure}
\centering
\includegraphics[scale=1]{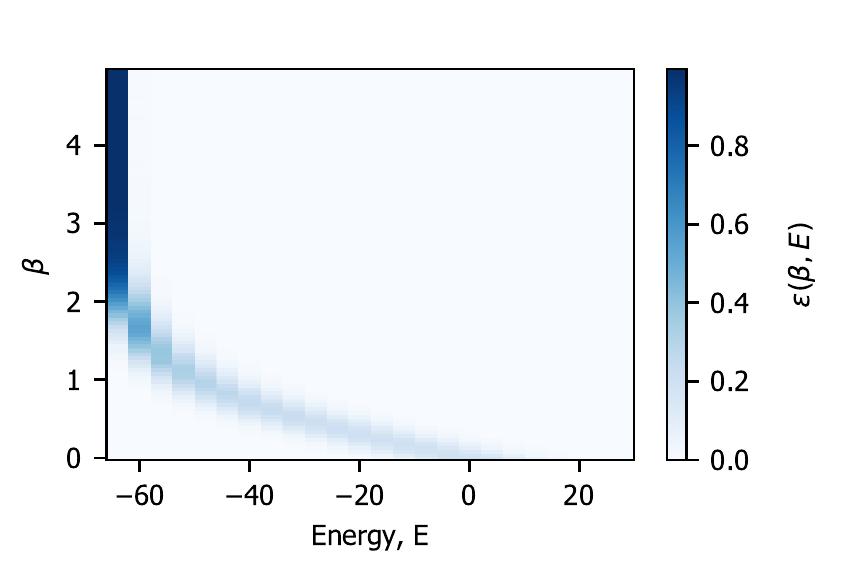}
\caption{Energy distribution $\epsilon (\beta, E)$ of the training set as a function of inverse temperature $\beta$ and energy $E$. The plot above has been generated for system size $N=8$.}
\label{fig:app_energy_dist}
\end{figure}

To independently confirm the neural network predictions, we can analyze whether we can identify the physics of what the network is learning and reproduce its predictions by another physical model. From the training set, we can construct a density of states distribution, $\epsilon$. In particular, the density of states can be written as a function of energy, $E$, and inverse temperature, $\beta$,
\begin{align}
\epsilon(\beta, E) = \frac{\sum_{n=1}^{N}\delta_{E,E_n}\delta_{\beta,\beta_n}}{\sum_{n=1}^N\delta_{\beta,\beta_n}}.
\label{eq:dens_dist}
\end{align}
Here, $\delta_{a,b}$ is the Kronecker-delta symbol ($\delta_{a,b}=1$ for $a=b$ and $\delta_{a,b}=0$ for $a\neq b$), $E_n$ ($\beta_n$) is energy (label) of the $n$-th configuration in the training set and $N$ is the number of configurations in the training set. We show the distribution $\epsilon$ obtained for the system size $N=8$ (128 spins) in Fig.~\ref{fig:app_energy_dist}.

We use the distribution \eqref{eq:dens_dist} to calculate the most likely $\beta =: \beta_{\ssm pred}$ for each configuration at a given energy, which immediately allows us to evaluate the relation between the assigned $\beta$ and $\beta_{\ssm pred}$. Using the density of states we are able to reproduce the behavior in Fig.~\ref{fig:divergence_igt} (see Appendix~\ref{app:igt}). We show the detailed calculation and the dependencies of the predicted $\beta_{\ssm pred}$ and its derivative $\mathcal D(\beta_{\ssm label})$ as a function of the true $\beta_{\ssm label}$ in Appendix \ref{app:igt}. This gives us a numerical evidence that the network is learning the density of states distribution shown in Fig.~\ref{fig:app_energy_dist}. We identify the logarithmic scaling (with system size) of the critical $\beta^*$ predicted from the density of states (shown in blue in Fig.~\ref{fig:scaling_igt}) analogously to the predictions obtained from the neural net model.

\begin{figure}
\centering
\includegraphics[scale=0.25]{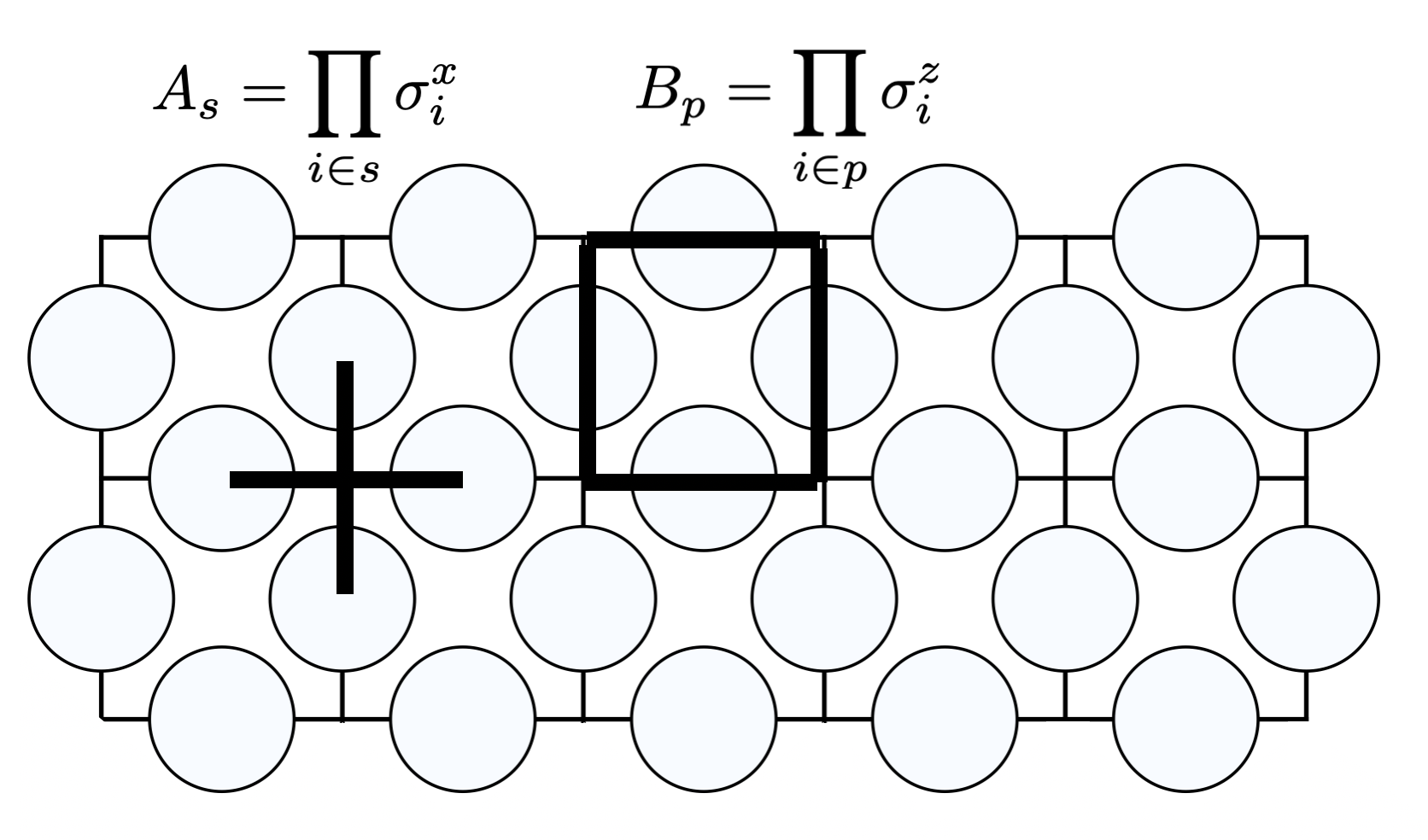}
\caption{Illustration of plaquette, $B_p$, and vertex, $A_s$, operators in the lattice.}
\label{fig:lattice_vertex_plaquette}
\end{figure}

\section{The toric codes and its generalizations}

So far we have analyzed the performance of our method on the cross-over of a {\em classical} spin-$1/2$ model. When going to {\em quantum} models, two complications arise, related to the {\em input} and {\em output} of our predictive model. For classical systems, simple spin configurations are the natural input. For quantum systems, generically entanglement in the form of non-classically correlated configurations plays a key role. Consequently, the choice of training data needs to either reflect some prior knowledge of the system, or one has to sample over various classical projections of the entangled wave function. On the output side, one can either target a finite-temperature transition, or investigate a quantum phase transition at zero temperature. In the former, the output of the predictive model stays the same: $\beta_{\ssm pred}$, the inverse temperature. For zero temperature transitions, one can still investigate a single-parameter family of Hamiltonians $H(\beta)$. The obvious prediction task is then to reproduce the tuning parameter $\beta$, rather then the temperature.

We now turn to a concrete model of a quantum phase transition in a system without a local order parameter. The obvious generalization of (\ref{eqn:igt}) is the application of a transverse field \cite{FradkinSusskind1978, Fradkin2013, Sachdev2018}
\begin{equation}
	\label{eqn:tr}
	H_{\ssm TR} = - \sum_p\prod_{i\in p}\sigma^z_i - g \sum_l\sigma^x_l.
\end{equation}
The model above is very well studied, has a confinement-deconfinement transition at a critical $g^*$, and is a working horse for the study of $\mathds Z_2$ spin-liquids. Instead of directly working with this simple model we go beyond (\ref{eqn:tr}) in two ways: (i) We restrict ourselves to a subset of gauge-invariant ground states by moving to the toric code \cite{Kitaev2006}. (ii) We generalize the transverse field to allow for an exact solution. We detail both steps in the following. 

The IGT of Eq. (\ref{eqn:tr}) has a local $\mathds Z_2$ gauge degree of freedom. The generators of this gauge transformation are the vertex operators
\begin{equation}
	A_s=\prod \limits_{i \in s} \sigma_i^x,
\end{equation}
that consist of a product of $\sigma_x$ operators along a vertex, $s$, of the lattice. The geometry of the vertex operator is illustrated in Fig.~\ref{fig:lattice_vertex_plaquette}. The operators $A_s$
commute with the Hamiltonian, i.e, $[H_{\ssm TR},A_s]=0$ for all vertices $s$. In other words, one can obtain an eigenstate by changing the sign of the classical $\sigma_z$-variables of another eigenstate, as long as one does so for all spins connected to one vertex. The toric code Hamiltonian
\begin{equation}
	H_{\ssm TC}= -\sum \limits_{s} A_s + H_{\ssm IGT} = -\sum \limits_{s} A_s  -  \sum \limits_{p}B_p,
\label{eq:toric}
\end{equation} 
elevates the generators of the gauge transformation to a term in the Hamiltonian. As a consequence, the ground states of the toric code correspond to the gauge-invariant ground states of $H_{\ssm TR}$ \cite{Fradkin2013}. For our numerical purposes below, we largely benefit from the exact solution of the above Hamiltonian: We can write one of the four (un-normalized) ground states as \cite{Kitaev2003}
\begin{align}
|{\rm TC}\rangle=\frac{1}{2}\prod \limits_{s} (1+B_p) |0_x\rangle,
\end{align}
where $|0_x\rangle$ is a reference state with all spins up in the $\sigma^x$ basis. Then, applying products of Pauli $z$-matrices along the two non-contractible loops yields the other three orthogonal ground states. We can easily see that the ground states are indeed gauge invariant by applying gauge transformations, obtaining $A_s|{\rm TC}\rangle=|{\rm TC}\rangle$.

Applying a transverse field a spin-model typically excludes an exact solution. The present case in no difference. However, in a recent publication, Chamon and Castelnovo introduced the following generalization of the toric code \cite{CastelnovoChamon2007, CastelnovoChamon2008, TsomokosOsborne2011, Valenti2019}
\begin{align}
H&=H_{\ssm TC}+\sum \limits_{s} e^{-\beta \sum \limits_{i \in p}\lambda_i\sigma_i^{x}}\nonumber \\ &=- \sum \limits_{s} A_s +  \sum \limits_{p}\left(-B_p+\sum \limits_{s} e^{-\beta \sum \limits_{i \in p}\lambda_i\sigma_i^{x}}\right), 
\label{eq:hamil_beta0}
\end{align} 
where $\lambda_i \in [-1,1]$ describes the particular configuration of added background fields and $\beta > 0$ characterizes their amplitude. A transition to a topologically trivial phase occurs at a critical value of the field strength $\beta_c$. The field configuration $\lambda_i$ influences the critical value $\beta_c$. A detailed analysis of this phase transition has been provided in \cite{Valenti2019}. 

To finish our discussion of these exactly solvable models we write the ground state of \eqref{eq:hamil_beta0}
\begin{align}
|\Psi\rangle &= \frac{1}{\sqrt{Z}}e^{\frac{\beta}{2}\sum \limits_{i} \lambda_i \sigma_i^x}|\rm TC\rangle \nonumber \\
&=\frac{1}{\sqrt{Z}}\sum \limits_{h\in H} e^{\frac{\beta}{2}\sum \limits_{i} \lambda_i \sigma_i^x(h)}h |0_x\rangle.\label{eq:toric_ground_mod}
\end{align}
This ground state is four-fold degenerate when periodic boundary conditions are considered \cite{Kitaev2003}. We denote with $H$ the abelian group whose elements $h$ are all possible operations defined by the action of products of plaquette operators on an initial (reference) spin-configuration $|0_x\rangle$. By $\sigma_i^x(h)$ we denote the eigenvalue of the operator $\sigma_i^x$ on the eigenstate $h |0_{x}\rangle$. As a consequence, the term $\sigma_i^x(h)$ can take the values $\pm 1$. The normalization factor, $Z$ corresponds to the partition function for this ground state and is given by 
\begin{align}
Z:=\sum_{h \in H} e^{\beta\sum \limits_{i} \lambda_i \sigma_i^x(h)}. \nonumber
\end{align}

\begin{figure}
\centering
\includegraphics[scale=1]{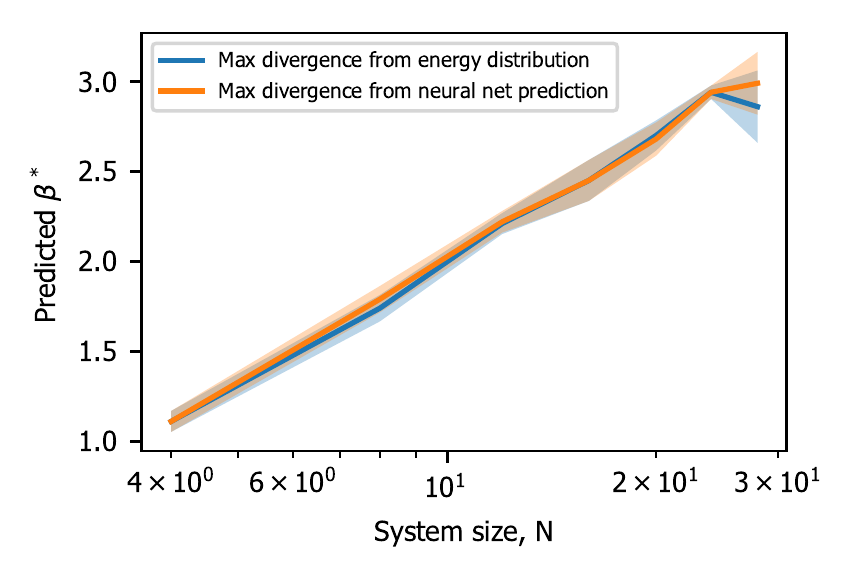}
\caption{Positions of critical $\beta^*$ as a function of a system size $N$. We show the scaling obtained from the unsupervised learning method and the scaling obtained from density of states in blue and orange respectively. The shaded areas represent the error bars. Error bars correspond to standard deviation from the mean $\tilde{\beta}^*$ evaluated by averaging over $\beta^*$ predicted by five separately trained neural nets.}
\label{fig:scaling_igt}
\end{figure}

With these considerations we are now in the position to show that the analysis of the predictive model can point out the topological phase transition of this quantum model as well. Unlike in the IGT, discussed in the previous section, the highly entangled ground states of the modified toric code model \eqref{eq:hamil_beta0} are not fully characterized by a spin configuration alone. On the other hand, Eq. \eqref{eq:toric_ground_mod} provides a closed analytical form for the ground states of the family of the Hamiltonians \eqref{eq:hamil_beta0}. In addition to that, these ground states are only four-fold degenerate in the topological phase. We take advantage of the knowledge of the modified toric code ground states and show this to be sufficient for identification of the phase transition from the predictive model.

\begin{figure}
\centering
\includegraphics[scale=1]{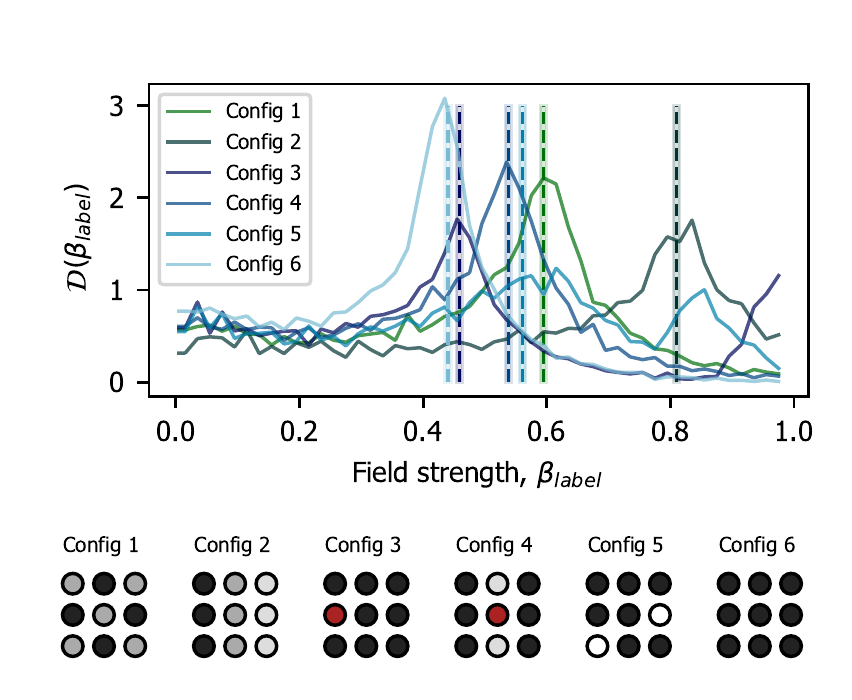}
\caption{Phase transition of the toric code model extracted from unsupervised learning on $\sigma_x$ configuration of the ground states. The derivative $\mathcal D(\beta_{\ssm label})$ of the prediction of neural network model designed to predict field strength $\beta$ is shown for six different field configurations. The configurations are illustrated in the bottom panel of the figure. Black circles denote field in positive direction (shades of grey denote strength), red circles denote field in the negative direction and empty circles denote no field present. The dashed vertical lines denote the position of the phase transition for a given field configuration evaluated using fidelity susceptibility.}
\label{fig:div_toric_sigmaz}
\end{figure}

\subsection{Projection onto spin configurations}

We consider a projection of the ground states of the Hamiltonian \eqref{eq:hamil_beta0} onto the $\sigma_x$ and $\sigma_z$ bases. These two types of projections correspond to experimentally accessible measurements and we show that both allow to detect the topological phase transition of the full quantum model. As for the IGT cross-over analyzed previously, we are yet again in the situation where we are able to input a configuration into the predictive model and ask it to predict a continuous parameter. 

There are two crucial differences here: First, we are considering a zero temperature topological phase transition that is driven by the applied field strength $\beta$. The second difference lies in the behavior of the projected spin configurations in the two phases. In particular, we are able to draw parallels to phase transitions of classical spin models. As we elaborate below, choosing a basis to project on corresponds to mapping the phases of the quantum model to phases of a specific classical spin model.

\subsubsection{The $\sigma_x$--projection}

Let us first consider the projection onto the $\sigma_x$ basis. We notice that the ground state \eqref{eq:toric_ground_mod} represents a superposition of $x$-spin-configurations $|S_h\rangle:=h|0\rangle_x$ for all elements of the group $H$. All states $|S_h\rangle$
fulfill the so-called closed loop condition
\begin{equation}
A_s|S_h\rangle=|S_h\rangle
\end{equation}
for all values of $\beta$. In connection to the IGT, this corresponds to the condition of gauge invariance. More concretely, local constraints are imposed, that the product of $\sigma_x$ eigenvalues around a vertex is equal to one. The value of the field strength, $\beta$ influences the weight of a given spin configuration (see Eq.~\eqref{eq:toric_ground_mod}). Therefore, the probability to obtain a particular configuration $|S_h\rangle$ after projection onto $\sigma_x$-basis is given by
\begin{align}
p(S_h)=|\langle S_h|\Psi(\beta)\rangle|^2=\frac{e^{\beta\sum \limits_{i} \lambda_i \sigma_i^x(h)}}{\sum \limits_{\tilde{h} \in H} e^{\beta\sum \limits_{i} \lambda_i \sigma_i^x(\tilde{h})}}.
\end{align}

We can understand the physics of the $\sigma_x$-projected ground state by first considering limiting cases of the field strength $\beta$. When $\beta\to 0$, the ground state \eqref{eq:toric_ground_mod} corresponds to the ground state of the pure toric code Hamiltonian \eqref{eq:toric}. Therefore, when projected onto the $\sigma_x$ basis, all possible $|S_h\rangle$ are equally likely (since all $|S_h\rangle$ are weighted equally in the full eigenstate). When $\beta\to\infty$, on the other hand, all configurations but $\ket{S}=\ket{0}$ are exponentially suppressed and hence, the projected spin configurations are always ordered. 

Thus, what used to be a topological phase transition of the full quantum state is now a transition from disordered spin-configurations ($\beta$ small) to an ordered spin-configuration ($\beta$ large, all spins up). We observe that, provided there is a finite $\beta$ at which the transition between ordered and disordered configurations manifests itself, we obtained a phase transition that shows resemblance to the phase transition of the $2D$ Ising model. We show that indeed the $2D$ Ising model and its phase transition can be recovered by a simple change of variables, see Appendix \ref{app:ising}.

Let us now explore the topological phase transition in the toric code model using the unsupervised learning method we introduced above. We train a neural network on the projected $\sigma_x$ configurations labeled with the field strength, $\beta$. We used a network consisting of two convolutional ($100$ filters, kernel size $3$ and $2$), one dense layer with $100$ neurons and one dropout layer with dropout rate $0.15$. We train the neural network on $59950$ configurations containing $100$ different values of $\beta$ between $0$ and $1$. All the simulations were performed for system size $N=8$.
Once the model is trained we apply it on $2000$ new configurations for $30$ different values of $\beta$ and evaluate the derivative $\mathcal D(\beta_{\ssm label})$ (\ref{eq:notdivergence}) of the outcome, see Fig.~\ref{fig:div_toric_sigmaz}. We show $\mathcal D(\beta_{\ssm label})$ for six example field configurations. As shown in \cite{Valenti2019} the position and the existence of the phase transition is strongly dependent on the distribution of added fields.


It was shown in \cite{CastelnovoChamon2008} that the topological phase transition of the generalized toric code model can be determined from the behavior of the fidelity between two ground states with slightly varied field strengths ($\delta\beta\to 0$)
\begin{equation}
F_{\beta}=\langle\Psi(\beta)|\Psi(\beta+\delta\beta)\rangle.
\label{eq:fidelity}
\end{equation}
In other words, we calculate the overlap of two ground state wave functions with applied fields whose magnitudes are very close to each other. We can indeed observe a change in the behavior of the overlap in the neighborhood of the phase transition. The rate of this change is better analyzed by studying the derivative of the quantity in Eq.~\eqref{eq:fidelity}, the so-called fidelity susceptibility
\begin{equation}
\chi_F=-\left.\frac{\partial^2\ln F}{\partial^2 (\delta\beta)}\right\rvert_{\delta\beta=0}.
\end{equation} 
We observe in Fig.~\ref{fig:div_toric_sigmaz} that the dashed lines determined from fidelity susceptibility calculation are in good agreement with the maximum of the peaks of the derivative $\mathcal D(\beta_{\ssm label})$ of the predictive model. We show details of the fidelity susceptibility calculation in Appendix \ref{app:ising}. 

\subsubsection{The $\sigma_z$--projection}

\begin{figure}
\centering
\includegraphics[scale=1]{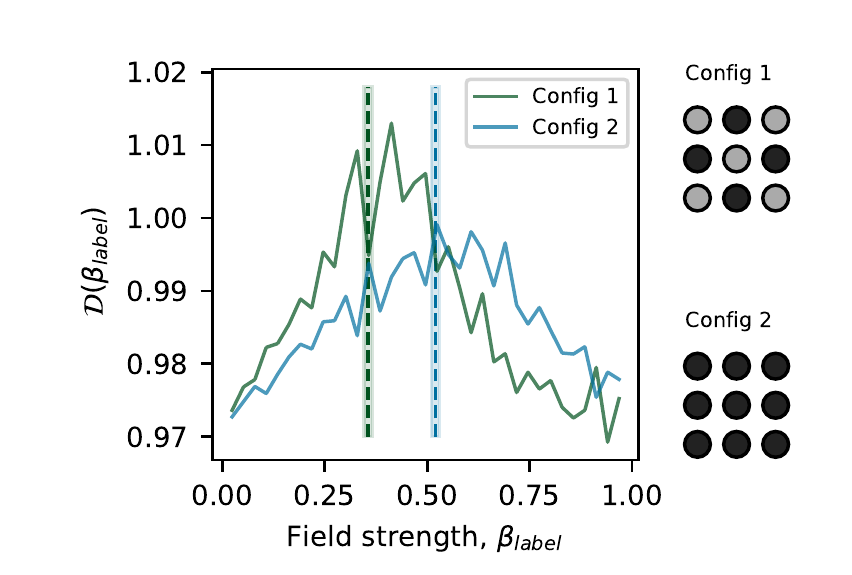}
\caption{Phase transition of the toric code model extracted from unsupervised learning on $\sigma_z$-projected configurations of the ground states. The derivative $\mathcal D(\beta_{\ssm label})$, of the prediction of the neural network model designed to predict the field strength $\beta$ is shown for $2$ different field configurations as a function of assigned $\beta_{label}$. The configurations are illustrated in the right hand panel of the figure. Black circles denote fields in positive direction (shades of grey denote weaker field strength).The dashed vertical lines correspond to the position of the phase transition determined using fidelity susceptibility.}
\label{fig:div_toric_sigmax}
\end{figure}

We can ask whether a particular projection is necessary to determine the topological phase transition from the spin configurations alone. Let us consider measuring the ground state in the $\sigma_z$ basis instead of $\sigma_x$. In order to simplify mathematical expressions let us without loss of generality choose a different state from the ground state manifold
\begin{align}
|\Psi_{z}\rangle=\frac{1}{\sqrt{Z_{z}}}e^{\frac{\beta}{2}\sum \limits_{i} \lambda_i \sigma_i^x}\sum \limits_{g\in G} g |0_z\rangle,
 \label{eq:toric_ground_mod_x} 
\end{align}
Here, analogously to  Eq.~\eqref{eq:toric_ground_mod}, $G$ is the abelian group of possible products of vertex operators and $\ket{0_z}$ is the reference state. Note that we chose a different reference state. As opposed to Eq.~\eqref{eq:toric_ground_mod}, all spins of the reference state are aligned in an eigenstate of $\sigma_z$ instead of $\sigma_x$. The normalization is denoted with $Z_{z}$ and not elaborated on further here.

Let us again examine the limiting behavior of $\beta$ if $\sigma_z$ was measured on every spin of the state \eqref{eq:toric_ground_mod_x}. If $\beta\to 0$ we obtain the exact toric code ground state. Projective measurement of $\sigma_z$ on this ground state then results in the configuration $S_g=g|0_z\rangle$, hence the closed loop (plaquette) conditions $B_p |S_g\rangle=|S_g\rangle$ are fulfilled. Every configuration $g|0_z\rangle$ fulfilling these constraints is obtained with equal probability. We note here, that the local plaquette constraints are in exact correspondence to the IGT local constraints fulfilled in the zero temperature phase.

Applying the same logic as in the case of the $\sigma_x$ projection, we can conclude that in the case $\beta\to\infty$ we arrive at a completely polarized state, where all spins are aligned in the $x$-direction. If we now project onto a $\sigma_z$ eigenstate, the vertex constraints will not stay preserved. In fact, any configuration in $\sigma_z$ basis will be obtained with equal probability. Hence, we find that in the $\sigma_z$ projection the phase transition arises from a quite different process than we observed before: for small $\beta$ the system would be in the state where loop conditions are preserved, while for large $\beta$ they are violated. While in the case of $\sigma_x$ the phase transition simply changes the weight for some states from the set preserving loop condition, in the case of $\sigma_z$ projection we transition from the state where all the states preserving loop condition are weighted equally to the phase where the loop constraints are completely violated. We can therefore draw parallels to the previosly examined IGT transition at finite temperature. In particular, in both cases we observe phases that can be distinguished by checking for a violation of the local closed loop constraints. However, there is a crucial difference between these two transitions. IGT exhibits a finite temperature cross-over and the violation of local constraints is a result of thermal excitations. Here, we consider a quantum phase transition at zero temperature, where the local constraints are violated due to the interplay with added perturbations. In particular, for IGT in the thermodynamic limit there is only a crossover at finite temperature whereas the quantum phase transition we consider here occurs at a finite field strength $\beta$ in the thermodynamic limit as well.

We employ the unsupervised learning technique on the $\sigma_z$ projection of the modified toric code ground state \eqref{eq:toric_ground_mod_x} with the strength of the background field $\beta$ as a label for the supervised part of the protocol. This time our neural net model consists of two convolutional layers (with $128$ filters and kernel size $2$) and three dense layers (with $100$, $100$ and $50$ neurons respectively).

We show the results for $N=4$ ($32$ spins) and two different field configurations in Fig.~\ref{fig:div_toric_sigmax}. The reduction of the system size and number of field configurations presented here are a consequence of constructing projections onto the  $\sigma_z$-basis from the ground state containing $\sigma_x$ fields: Mixed $\sigma_z\sigma_x$ terms make Monte Carlo update computationally significantly more expensive (for details see Appendix~\ref{app:ising}).

\subsection{Phase transition determination from the stabilizer expectation values}

Finally, we discuss on how to obtain the topological phase transition in the toric code model by extracting necessary information by measurements that can be readily performed on the quantum state at hand and do not require projections onto the spin configurations. It was shown in \cite{Valenti2019} that the behavior of the expectation values of the stabilizer operators are intimately related to the position of the topological phase transition in the toric code model. We use our predictive model to evaluate the position of the phase transition from the expectation value of the stabilizer operators to offer an alternative method to determine the position of the phase transition.

\begin{figure}
\centering
\includegraphics[scale=1]{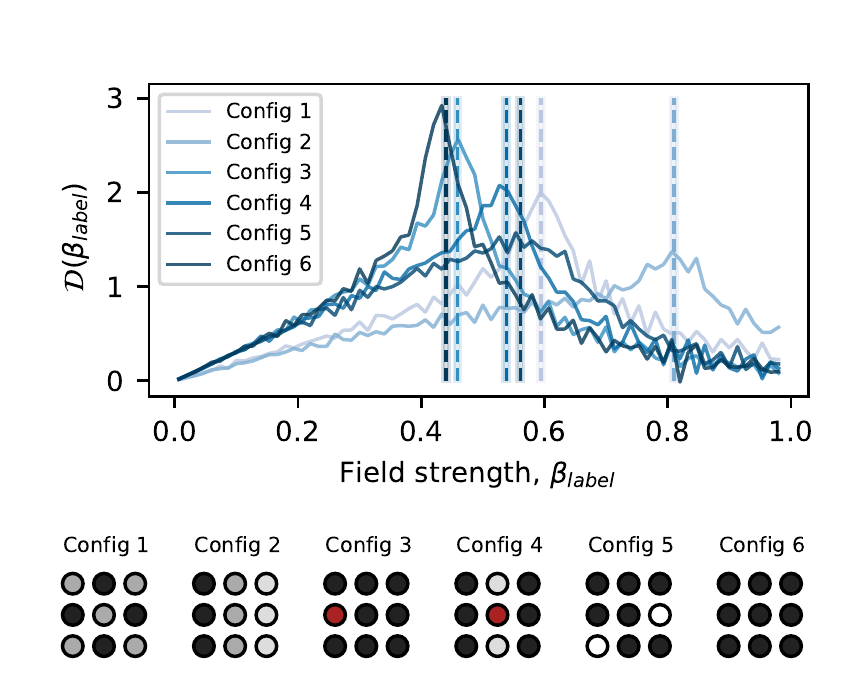}
\caption{Phase transition of the toric code model extracted from unsupervised learning on expectation values of a stabilizer operator, $\langle B_p\rangle$, in the ground state. The derivative $\mathcal D(\beta_{\ssm label})$ of the prediction of the neural network model designed to predict the field strength $\beta$ is shown for six example field configurations. Black circles denote fields in positive direction (shades of grey denote a weaker field strength), red circles denote fields in negative direction and empty circles denote no field present.}
\label{fig:div_toric_exp}
\end{figure}

As in the previous sections, we train a neural network to predict the value of the field strength amplitude, $\beta$. This time we use as an input the expectation value of the plaquette operator, $\langle B_p\rangle$ (with $\beta$ as a label). Then we use the network to predict the field strength $\beta$ for the expectation values of $B_p$ evaluated with respect to the new set of quantum states. We use a neural network with two dense layers (with $20$ neurons each). The derivative $\mathcal D(\beta_{\ssm label})$ of the predictive model is shown as a function of field strength for six distinctive field configurations is shown in Fig.~\ref{fig:div_toric_exp}. We again compare to the position of the phase transition obtained by fidelity susceptibility method (dashed lines) and observe an excellent agreement.

While the connection between expectation values of stabilizer operators and the position of the phase transition have not been shown analytically, another numerical evidence was provided in \cite{Nielsen2019}. The authors examine direct detection of anyons, a process that can be mapped onto the expectation values which we investigated here. The presence of anyons is then immediately tied to the existence of topological order. We elaborate on the connection to the present work in Appendix~\ref{app:ising}.

\section{Discussion}

Unsupervised machine learning techniques for phase classification in condensed matter physics are potentially powerful tools for the discovery of new quantum phases. Due to the lack of local order parameters, phases exhibiting topological order present a challenging task for unsupervised methods. In this work, we have shown that a novel unsupervised method, namely the analysis of predictive neural network models \cite{Schafer2019}, can reliably detect the violations of topological order, or a topological phase transition should it exist.

Topologically ordered states have been particularly challenging for unsupervised learning techniques, because the quantity characterizing topological order is inherently non-local and hard to identify from raw data. In the method presented here, we trained the network on an arbitrary continuous parameter associated to the state and then analyzed the errors in the network predictions. We presented numerical evidence that these prediction errors are signatures of a phase transition. We showed that this conclusion was independent of the particular type of phase transition present in the system and the type of the input data.

Providing the resolution to the problem of finding the cross-over temperature in the IGT and its generalizations in an unsupervised manner is the first step towards developing reliable techniques that can be applied to study the models whose phase diagrams are not yet fully understood.

\section*{Acknowledgements}
We thank Juan Carrasquilla for fruitful discussions. We acknowledge Mark H. Fischer for contributions to early versions of the code. We are grateful for financial support from the Swiss National Science Foundation, the NCCR QSIT. This work has received funding from the European Research Council under grant agreement no. 771503.

\appendix

\section{IGT: Predictive model}
\label{app:igt}

We created the samples used for training (like those shown in Fig.~\ref{fig:igt_example}) of our model using Monte Carlo simulations. We created data for system sizes $N\times N\times 2$ with $N\in\{4,8,12,16,20,24,28\}$. For each system size we created $100$ different values of $\beta\in[0,5]$. We generated $20 000$ configurations for each pair $[\beta$, N]. The neural net we used consists of $2$ convolutional ($128$ filters, kernel size $3$) and $2$ dense layers ($300$ and $100$ neurons respectively). We trained the network by minimizing the mean-squared-error loss function
\begin{equation}
L^{\ssm mse}(\beta_{\ssm pred}-\beta_{\ssm label})=\frac{1}{n}\sum_n(\beta_{\ssm pred}-\beta_{\ssm label})^2,
\end{equation}
where $\beta_{\ssm pred}$ is the $\beta$ determined by the network and $\beta_{\ssm label}$ is the label of the given input sample, $n$ is the batch size. The predictions of $\beta$ by the network and their divergences are shown in Fig.~\ref{fig:prediction_igt} and Fig.~\ref{fig:divergence_igt}.

In order to evaluate the error bars of the neural net predictions, we repeated the training procedure outlined above for $5$ separate models (identical construction, separately generated training sets). Then we evaluated standard deviation of the critical $\beta^*$.

We can replicate the predictions achieved by a neural network using a density of states based model as explained in the main text. Let us consider lattice configurations (training samples) $X_n$ with their assigned inverse temperature labels $\beta_n=\beta_{label}(X_n)$. We can evaluate an energy, $E_n$ of each of these configurations using the formula
\begin{equation}
E = -J \sum_p \prod_{i\in p}\sigma_i^z,
\label{eq:app_energy}
\end{equation} 
where the first summation is over all plaquettes, $p$, whereas the second summation is over spins within each plaquette. For convenience we choose $J = 1$. Then we can construct the density of states distribution of the training set
\begin{align}
\epsilon(\beta, E) = \frac{\sum_{n=1}^{N}\delta_{E,E_n}\delta_{\beta,\beta_n}}{\sum_{n=1}^N\delta_{\beta,\beta_n}}.
\end{align}
Here, $\delta_{a,b}$ is the Kronecker delta symbol ($\delta_{a,b}=1$ for $a=b$ and $\delta_{a,b}=0$ for $a\neq b$), $E_n$ is energy for the configuration $X_n$ evaluated using formula \ref{eq:app_energy} and $N$ is number of configurations $X_n$ in the training set.
We can write the energy distribution in the form above because the energy of the lattice configuration, $E$ is discrete by construction and $\beta$ is discretized in steps as explained above. An example of this distribution is shown in Fig.~\ref{fig:app_energy_dist} for the system size $N=8$. 

\begin{figure}
\centering
\includegraphics[scale=1]{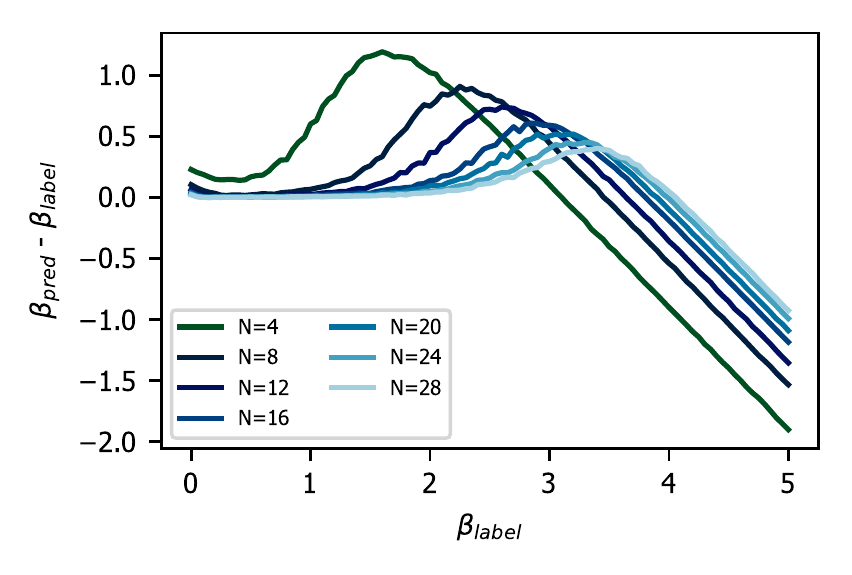}
\caption{Density of states based prediction of $\beta$. We plot the difference between true and predicted $\beta$, $\beta_{\ssm pred}-\beta_{\ssm label}$, as a function of $\beta_{\ssm label}$.}
\label{fig:app_beta_pred}
\end{figure}

\begin{figure}
\centering
\includegraphics[scale=1]{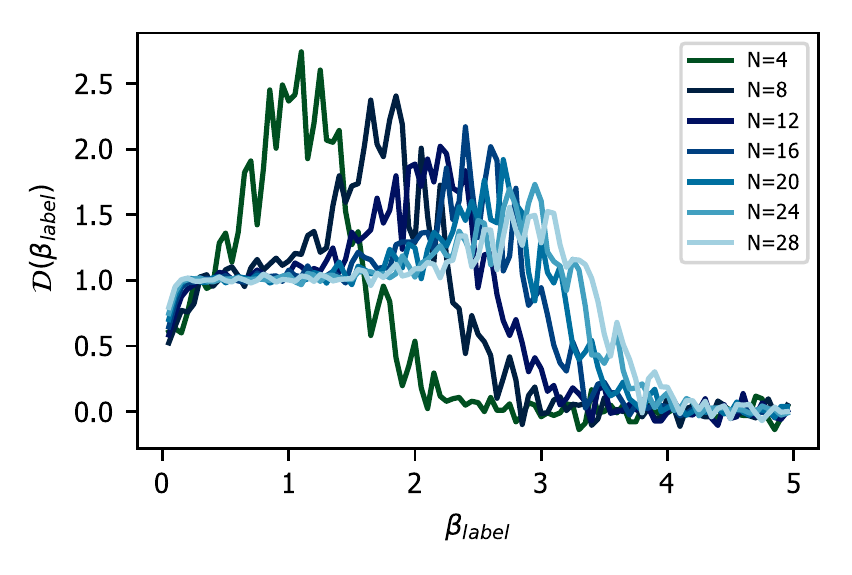}
\caption{We show the derivative of the density of states based prediction, $\mathcal{D}(\beta_{\ssm label})$, as a function of $\beta_{\ssm label}$.}
\label{fig:app_div_beta_pred}
\end{figure}

Having access to the energy $E_n$ of a given configuration $X_n$, we can then evaluate the average $\beta$ of all states with energy $E$, which we denote by $\beta^{av}$ for a configuration $X_n$ in the training set
\begin{align}
\beta^{av}(E)=\frac{\sum_{n=1}^N\delta_{E,E_n}\beta_n}{\sum_{n=1}^N\delta_{E,E_n}}.
\label{eq:app_beta_average}
\end{align}
The function above predicts the value of $\beta$ which is most likely for a given energy, $E$, given the energy distribution of the training set. We can use the function \eqref{eq:app_beta_average} to determine the relation between assigned labels, $\beta_n$ and values of $\beta$ predicted by our model
\begin{equation}
\beta^{est}(\beta) = \frac{\sum_{m=1}^{M}\beta^{av}(E_m)\delta_{\beta,\beta_m}}{\sum_{m=1}^M\delta_{\beta,\beta_m}},
\label{eq:app_beta_est}
\end{equation}
where $M$ is the number of configurations $X_m$ in an arbitrarily chosen test set. Using equation \eqref{eq:app_beta_est} we can predict the estimated $\beta$ for a range of true labels. In Fig.~\ref{fig:app_beta_pred} we show the difference between true and predicted $\beta$ as a function of true $\beta$. In Fig.~\ref{fig:app_div_beta_pred} we show the derivative of the estimated $\beta$ as a function of true $\beta$. Comparing with Fig.~\ref{fig:divergence_igt} we see that our model based on the density of states in the training set is reproducing well the actions of the neural net model we introduced in the main text. We have used maxima determined by the density of states as a dashed-line reference for the position of the transition in the main text.

\section{Toric code: Predictive model}
\label{app:ising}
\subsection{Mapping to Ising model}
The projection of the modified toric code ground state on the $\sigma_x$ basis can be understood by mapping to a classical Ising model.
Let us examine the ground state of the toric code with fields (\ref{eq:toric_ground_mod})
\begin{align}
|\Psi\rangle = \frac{1}{\sqrt{Z}}\sum \limits_{h\in H} e^{\frac{\beta}{2}\sum \limits_{i} \lambda_i \sigma_i^x(h)}h |0_x\rangle.\label{eq:toric_ground_mod_app}
\end{align}

We perform a projection of $|\Psi\rangle$ onto the $\sigma_x$ basis. As stated in the main text, the outcome of the projection are the configurations $|S_h\rangle$ fulfilling the closed-loop condition $A_s|S_h\rangle=|S_h\rangle$. In addition, the probability to obtain $|S_h\rangle$ after projection is given by 
\begin{align}
p(S_h)=|\langle S_h|\Psi\rangle|^2=\frac{e^{\beta\sum \limits_{i} \lambda_i \sigma_i^x(h)}}{\sum \limits_{\tilde{h} \in H} e^{\beta\sum \limits_{i} \lambda_i \sigma_i^x(\tilde{h})}}.
\label{eq:prob}
\end{align}

\begin{figure}
\centering
\includegraphics[scale=1]{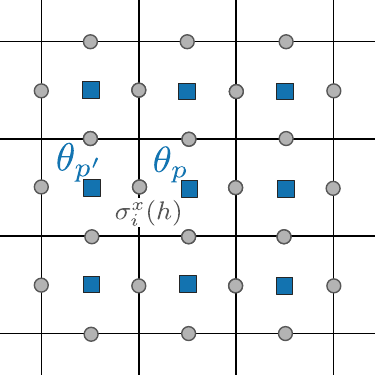}
\caption{Illustration of the mapping to a classical Ising model. The introduced pseudo-spins on the vertices are drawn in blue, the original spins in grey. The relation $\sigma_i^x(h)=\theta_{p}\theta_{p'}$ holds.}
\label{fig:Ising_mapping}
\end{figure}

The system can be mapped to the classical Ising model as follows \cite{CastelnovoChamon2008}. First we notice, that every group element $h$ uniquely determines the configuration $|S_h\rangle$ by applying $h$ to a reference spin configuration, which we choose to be $|0_x\rangle$ (all spins up in $x$-basis). Then, $|S_h\rangle=h|0_x\rangle$. In addition, $h$ corresponds to the product of a set $I_h$ of plaquette operators $h=\prod_{p \in {I_h}} B_p$. Every such set of plaquette operators corresponding to a spin configuration $|S_h\rangle$ can be mapped to the following pseudo-spin configuration: Artificial degrees of freedom (pseudo-spins) $\theta_p \in \{-1,1\}$ are introduced on every plaquette. The value of the pseudo-spin $\theta_p$ is determined by $I_h$: if $B_p \in I_h$ (plaquette flipped) it is equal to $-1$, else it is equal to one. As a consequence, the original spin-configuration $\{\sigma_i^x(h)\}$ (corresponding to $|S_h\rangle$) can be deduced from the pseudo-spin configuration $\theta_p$ by applying the rule $\sigma_i^x(h)=\theta_{p}\theta_{p'}$. Here, $p$ and $p'$ are the two adjacent plaquettes to spin $i$.
The geometry of the mapping is illustrated in Fig.~\ref{fig:Ising_mapping}.

Let us translate the mapping into the calculation of the probability $p(S_h)$. Inserting the rule $\sigma_i^x(h)=\theta_{p}\theta_{p'}$ to Eq. (\ref{eq:prob}) yields
\begin{align}
p(S_h)=p(\{\theta^h\})=\frac{e^{\beta\sum \limits_{\langle p,p' \rangle} J_{p,p'} \theta^h_p \theta^h_{p'}}}{\sum \limits_{\{\theta\}} e^{\beta\sum \limits_{\langle p,p' \rangle} J_{p,p'} \theta_p \theta_{p'}}},
\label{eq:prob_Sg}
\end{align}
with $J_{p,p'}=\lambda_i$ for the plaquettes $p,p'$ adjacent to edge $i$
and summing over nearest-neighbors $\langle p,p' \rangle$. The pseudo-spin configuration obtained from the group element $h$ by applying the explained mapping is denoted by the parameters $\{\theta^h\}$. In contrast, the sum over $\{\theta\}$ represents a sum over all possible pseudo-spin configurations.
We recognize the expression (\ref{eq:prob_Sg}) as Boltzmann weight for an Ising model with bond strengths $J_{p,p'}$ at temperature $T=1/({k_B \beta})$. The topological phase transition undergone by the studied perturbed toric code model hence shows the behavior of an Ising phase transition from disordered pseudo-spin configurations to ordered spin configurations after projecting onto the $\sigma_x$ basis.

\subsection{Calculation of the fidelity susceptibility}
\begin{figure}
\centering
\includegraphics[scale=1]{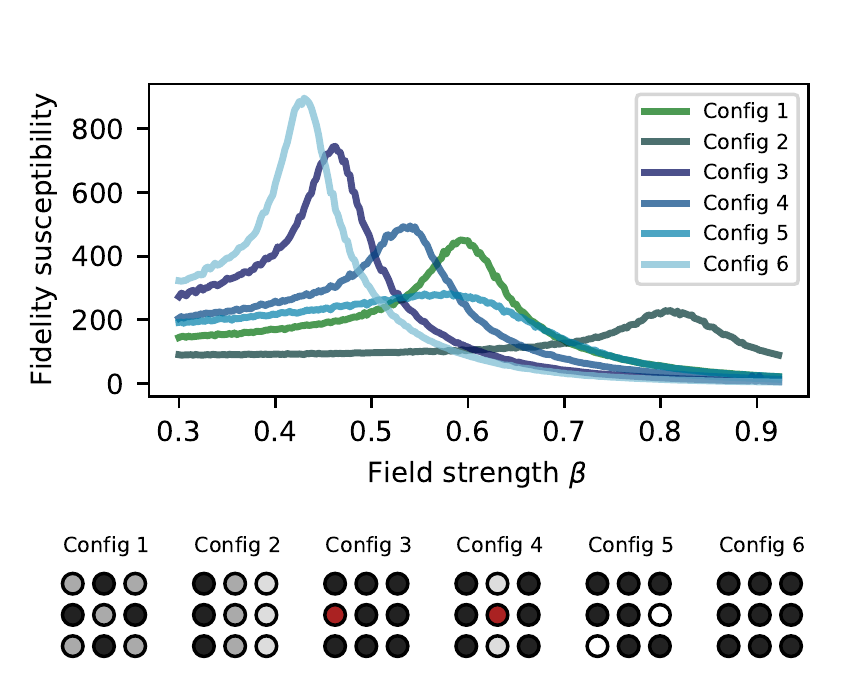}
\caption{Fidelity susceptibility numerically calculated for a lattice of length $L=20$. The different field configurations are illustrated in the exemplary configuration plots.}
\label{fig:fidelities}
\end{figure}

We compare the position of the phase transition found by the neural network to the transition indicated by the fidelity susceptibility \cite{YouLi2007}.
The fidelity susceptibility is defined as
\begin{align}
\chi_F=-\frac{\partial^2 \ln \langle \rm \Psi (\beta) |\rm \Psi (\beta+\Delta\beta) \rangle}{\partial (\Delta\beta)^2}\bigg|_{\Delta\beta =0},
\label{eq::fidsusceptibility}
\end{align}
where the state $|\Psi (\beta)\rangle$ is a ground state of a given Hamiltonian with respect to the parameter $\beta$. For our particular model, $|\Psi (\beta)\rangle$ is given in Eq. (\ref{eq:toric_ground_mod}).
It has been shown, that a divergence or maximum of the fidelity susceptibility $\chi_F$ indicates a second-order symmetry-breaking quantum phase transition \cite{VenutiZanardi2007, GuLin2009, ZanardiPaunkovic2006}. Numerical evidence suggests that topological phase transitions are indicated in the same way \cite{AbastoHamma2008}. We can calculate the fidelity susceptibility for the introduced disordered toric model as
\begin{align}
&\chi_F=\frac{1}{4}\frac{\sum \limits_{h \in H}(\sum \limits_{i} \lambda_i \sigma_i^x(h))^2 e^{\beta\sum \limits_{i}\lambda_i\sigma_i^x(h)}\cdot Z}{Z^2} \\
&\ \ \ \ \ \ -\frac{1}{4}\frac{(\sum \limits_{h \in H}(\sum \limits_{i} \lambda_i \sigma_i^x(h)) e^{\beta\sum \limits_{i}\lambda_i\sigma_i^x(h)})^2}{Z^2}.
\end{align}
We numerically evaluate the expression via Monte Carlo sampling for the different field configurations examined throughout this work and compare the position of the maximum with the position of the phase transition found by the neural network. The numerical simulation on a lattice of length $L=20$ of the fidelity susceptibility for different field configurations is shown in Fig.~\ref{fig:fidelities}.
The same analysis has been repeated for $L=4$.

The fidelity susceptibility can be connected to the heat capacity of the classical Ising model explained in the previous subsection, as elaborated in \cite{AbastoHamma2008} and \cite{Valenti2019}.

\subsection{Numerical simulation of projections}
The projection of the ground state of the perturbed toric model onto the $\sigma_x$ or $\sigma_z$ basis is in both cases simulated via Monte Carlo sampling. To project on the $\sigma_x$ basis, we aim to obtain a configuration $S_h$ sampled from the probability distribution $p(S_h)$ (\ref{eq:prob_Sg}). Such a configuration is reached via a Markov chain. More concretely, we start with a lattice with all spins up in $x$ basis and construct the Markov chain as follows: in each step (with given spin configuration $S_{h_i}$), a random plaquette $p$ is picked. The decision, whether the plaquette should be flipped and $S_{h_i}\to S_{h_{i+1}}$ is made via a Metropolis-Hastings test. The four spins around the chosen plaquette are flipped with probability 
\begin{align}
\frac{p(S_{h_{i+1}})}{p(S_{h_i})}=e^{\beta\sum \limits_{i \in p} \lambda_i (\sigma_i^x(h_{i+1})-\sigma_i^x(h_{i}))}.
\end{align}
After thermalization time, the spin configuration $S_h$ is obtained with probability $p(S_h)$ and a projection is simulated.

Projecting on the $\sigma_z$ basis follows the same principle with the caveat that the spin-flip probability is computationally expensive to calculate. We start from a state in the ground state manifold
\begin{align}
|\Psi_{z}\rangle &=\frac{1}{\sqrt{Z_{z}}}e^{\frac{\beta}{2}\sum \limits_{i} \lambda_i \sigma_i^x}\sum \limits_{g\in G} g |0_z\rangle. 
\end{align}
and project on the $\sigma_z$ basis.
In particular, let us examine the probability to obtain the spin configuration 
\begin{align}
|z_M\rangle=\prod \limits_{i \in M}\sigma_i^x |0_z\rangle,
\end{align}
where $M$ is a set of spins that are flipped in the configuration $|z_M\rangle$ with respect to the initial state $|0_z\rangle$. Then, the probability to project on $|z_M\rangle$ is given by
\begin{align}
&p(z_M)=|\langle z_M |\Psi_{z}\rangle|^2 \nonumber\\ 
&=\bigg(\frac{\sum \limits_{C} \prod \limits_{j\notin C_M} \cosh (\frac{\beta}{2}\lambda_j) \prod \limits_{i\in C_M} \sinh (\frac{\beta}{2}\lambda_j)   }{ \sum \limits_{C} \prod \limits_{j\notin C} \cosh (\beta\lambda_j) \prod \limits_{i\in C} \sinh (\beta\lambda_j)  }\bigg)^2,
\label{eq:prob_x}
\end{align}
where the closed loops $C$ correspond to the set of spins that are flipped when applying a product of vertex operators to the intial state $\prod A_s |0_z\rangle= \prod_{i \in C}\sigma_i^x |0_z\rangle$. The sum is over all possible closed loops, hence over all possible products of vertex operators. Similarly, the set $C_M$ can be constructed from the closed loop $C$ by flipping the spins in $M$
\begin{align}
\prod \limits_{i \in M}\sigma_i^x  \prod_{i \in C}\sigma_i^x |0_z\rangle= \prod_{i \in C_M}\sigma_i^x |0_z\rangle.
\end{align}
In order to obtain a spin configuration sampled from the distribution $p$ defined in Eq. (\ref{eq:prob_x}), we construct a Markov chain by starting with a spin configuration $|0_z\rangle$ in the $\sigma_z$ basis. In each step, a random spin is chosen and flipped (updating the spin configuration $|z_i\rangle$ to $|z_{i+1}\rangle$) with probability $p(z_{i+1})/p(z_{i})$.
As the computation of the spin flip probability is expensive, we simulate $z$-projections only for $2\times 4\times 4=32$ spins.

The neural network for the predictive model on the lattice with length $L=4$ (outcomes shown in Fig.~\ref{fig:div_toric_sigmax}) consists of two convolutional layers with $128$ neurons each and three dense layers with $100$, $100$ and $50$ neurons. Training was conducted on a set of 157960 examples in total, $144$ values of $\beta$ between $0$ and $1$. For evaluation of the trained model to predict the field parameter, the values for $\beta$ were chosen to be $72$ discrete steps between $0$ and $1$. A total of 100800 evaluation examples was generated, data augmentation (rotations, translations, mirror) led to an additional factor of $100$.

\subsection{Detection of Quasiparticles}
We elaborated in the main text, that measuring a stabilizer expectation value contains sufficient information to indicate the position of the topological phase transition. This behaviour can be related to a detection of the topological phase transition by measuring quasiparticles. More concretely, numerical evidence has been presented in \cite{Nielsen2019}, that a topological phase transition can be indicated by a detection of quasiparticles.
For the toric model, the toric Hamiltonian can be modified such that the ground state contains a pair of quasiparticles
\begin{align}
H_m=-\sum \limits_{p\neq p1, p2} B_p -  \sum \limits_{s}A_s +\sum \limits_{p= p1, p2} B_p.
\end{align}
At the plaquettes $p_1$ and $p_2$, the expectation value $\langle B_p \rangle_m=-1$ measured on the ground state shows the existence of a quasiparticle. Here, the subscript $m$ denotes that the expectation value is taken with respect to the ground state of the modified toric code. When adding a phase-transition driving perturbation paramatrized by a field $\beta$, the position of the phase transition is indicated by a divergence in the derivative $\partial_{\beta}\langle B_p \rangle_m$ with $p\in \{p_1,p_2\}$.
If the added perturbation is of the form 
\begin{align}
H_m \to H_m+\sum \limits_p e^{-\beta\sum \limits_{i \in p} \lambda_i \sigma_i^x},
\end{align}
the following relation holds
\begin{align}
\langle B_p \rangle=-\langle B_{p1} \rangle_m.
\end{align}
Here, the expectation value $\langle B_p \rangle$ is evaluated on the ground state of the model without quasiparticles (\ref{eq:toric_ground_mod}) examined throughout this work.
We conclude, that the divergence in the derivative of $\langle B_p \rangle$ indicates the position of the phase transition. We therefore understand, that the predictive model is able to reconstruct this behaviour as the accuracy of the predicted field strength depends on the slope of the expectation value $\langle B_p \rangle$.

\bibliographystyle{unsrt}
\bibliography{phtr_bib.bib}

\end{document}